\documentclass[12pt,oneside]{article}

\usepackage{amsmath}
\usepackage{graphicx}
\usepackage{epsfig}

\def\etal{{\it et al.\ }}
\newcommand{\ltwid}{\mathrel{\raise.3ex\hbox{$<$\kern-.75em\lower1ex\hbox{$\sim$}}}}
\newcommand{\gtwid}{\mathrel{\raise.3ex\hbox{$>$\kern-.75em\lower1ex\hbox{$\sim$}}}}
\newcommand{\flbox}[1]{\framebox(14,6){\parbox{12\unitlength}{\begin{center}
 #1 \end{center}}}}
\newcommand{\flboxa}[1]{\framebox(12,6){\parbox{12\unitlength}{\begin{center}
 #1 \end{center}}}}

\newcommand{\dln}{\line(0,1){4}}
\newcommand{\hln}{\line(1,0){10}}

\begin{document}
\thispagestyle{empty}
\setcounter{page}{0}
\renewcommand{\theequation}{\thesection.\arabic{equation}}

\vspace{2cm}

\begin{center}
{\bf\Large Generalizing Quantum Mechanics\\
for Quantum Spacetime}\footnote{To appear in the proceedings of the 23$^{\rm rd}$ Solvay
Conference, {\sl The Quantum Structure of Space and Time}, 12/1--3/05, Brussels}

\vspace{1.4cm}

James B.~Hartle

\vspace{.2cm}

{\em Department of Physics, University of California} \\
{\em Santa Barbara, CA 93106-9530 USA } \\
\end{center}

\vspace{-.1cm}

\centerline{{\tt hartle@physics.ucsb.edu}}
\vspace{1cm}
\centerline{ABSTRACT}

\vspace{- 4 mm}  

\begin{quote}\small
Familiar textbook quantum mechanics assumes a fixed background spacetime to define states
on spacelike surfaces and their unitary evolution between them.  Quantum theory has
changed as our conceptions of space and time have evolved.  But quantum mechanics needs
to be generalized further for quantum gravity where spacetime geometry is fluctuating and
without definite value. This paper reviews a fully four-dimensional, sum-over-histories,
generalized quantum mechanics of cosmological spacetime geometry. This generalization is
constructed within the framework of generalized quantum theory. This is a minimal set
of principles for quantum theory abstracted from the modern quantum mechanics of closed
systems, most generally the universe. In this generalization,
states of fields on spacelike surfaces and their unitary evolution are emergent properties
appropriate when spacetime geometry behaves approximately
classically. The principles of generalized quantum theory allow for the further
generalization that would be necessary were spacetime not fundamental. Emergent spacetime
phenomena are discussed in general and illustrated with the example of the classical
spacetime geometries with large spacelike surfaces that emerge from the `no-boundary'
wave function of the universe. These must be Lorentzian with one, and only one, time
direction. The essay concludes by raising the question of whether quantum mechanics
itself is emergent.
\end{quote}
\baselineskip18pt
\noindent

\vspace{5mm}

\newpage 
\setlength{\baselineskip}{.2in}
\setcounter{equation}{0}
\setcounter{footnote}{0}
\section{Introduction}

Does quantum mechanics apply to spacetime? This is the question the organizers asked me to 
address.  It is an old issue.  The renowned Belgian physicist L\'eon Rosenfeld
wrote one of the first papers on quantum gravity \cite{Ros30}, but late in his
career came to the conclusion that the quantization of the gravitational field would be
meaningless\footnote{Rosenfeld considered the example of classical geometry curved by the
expected value of the stress-energy of quantum fields. Some of the difficulties with this
proposal, including experimental inconsistencies, are discussed by Page and Geilker
\cite{PG81}.} \cite{Ros63,Ros66}. Today, there are probably more colleagues of the opinion
that quantum theory needs to be replaced than there are who think that it doesn't apply to
spacetime. But in the end this is an experimental question as Rosenfeld stressed .

This lecture will answer the question as follows: {\sl Quantum
mechanics {\it can} be applied to spacetime provided that the usual textbook formulation
of quantum theory is suitably generalized.} A generalization is necessary because, in one
way or another, the usual formulations rely on a fixed spacetime geometry to define states
on spacelike surfaces and the time in which they evolve unitarily one surface to another.
But in a quantum theory of gravity, spacetime geometry is generally fluctuating and
without definite value.  The usual formulations are emergent from a more general
perspective when geometry is approximately classical and can supply
the requisite fixed notions of space and time.

A framework for investigating generalizations of usual quantum mechanics can be abstracted
from the modern quantum mechanics of closed systems \cite{Gri02, Omn94, Gel94} which
enables quantum mechanics to be applied to cosmology. The resulting framework ---
generalized quantum theory \cite{Har91a, Har95c, Ishsum95} --- defines a broad class of
generalizations of usual quantum mechanics.

A generalized quantum theory of a physical system (most generally the universe) is built
on three elements which can be very crudely characterized as follows:

\begin{itemize}

\item The possible fine-grained descriptions of the system.
\item The coarse-grained descriptions constructed from the fine-grained ones.
\item A measure of the quantum interference between different coarse-grained descriptions
incorporating the principle of superposition

\end{itemize}
We will define these elements more precisely in Section 6, explain how they are used to
predict probabilities, and provide examples. But, in the meantime, the two-slit experiment
shown in Figure 1 provides an immediate, concrete illustration.

\begin{figure}[t]
\begin{center}
\includegraphics[width=10cm]{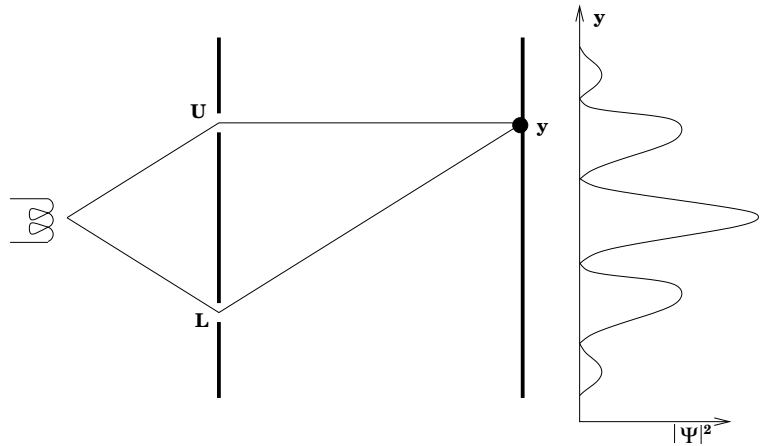}
\caption{\small The two-slit experiment. An electron gun at left emits an electron
traveling towards a screen with two slits, $U$ and $L$, its progress in space recapitulating
its evolution in time. The electron is detected at a further screen in a small
interval $\Delta$ about the position $y$. It is not possible to
assign probabilites to the alternative histories of the electron in which it
went through the upper slit $U$ on the way to $y$, or through the lower slit $L$
on the way to $y$ because of the quantum interference between these two histories.}
\label{fig:twoslit}
\end{center}
\end{figure}

A set of possible fine-grained descriptions of an electron moving through the two-slit
apparatus are its Feynman paths in time (histories) from the source to the detecting
screen.  One coarse-grained description is by which slit the electron went through on its way to detection in an interval $\Delta$ about a position $y$ on the screen at a later time.  Amplitudes $\psi_U(y)$ and $\psi_L(y)$ for the two coarse-grained
histories where the electron goes through the upper or lower slit and arrives at a point $y$ on
the screen can be computed as a sum over paths in the usual way (Section 4). 
The natural measure of interference
between these two histories is the overlap of these two amplitudes integrated over the interval
$\Delta$ in which the electron is detected. In this way usual quantum mechanics is a special case
of generalized quantum theory.  

Probabilities cannot be assigned to the two coarse-grained histories illustrated in Figure 1
because they interfere. The probability to arrive at $y$ should be the sum of the probabilities to
go by way of the upper or lower slit. But in quantum theory, probabilities are squares of
amplitudes and 
\begin{equation}
|\psi_U (y) + \psi_L (y)|^2 \not= |\psi_U (y) |^2 + |\psi_L (y)|^2\, .
\end{equation}
Probabilities can only be predicted for sets of alternative coarse-grained histories for which the 
quantum interference is negligible between every pair of coarse-grained histories in the set 
(decoherence).

Usual quantum mechanics is not the only way of implementing the three elements of generalized
quantum theory. Section 7 sketches a sum-over-histories generalized quantum theory of spacetime.
The fine-grained histories are the set of four-dimensional cosmological spacetimes with matter
fields on them. A coarse graining is a partition of this set into (diffeomorphism invariant)
classes. A natural measure of interference is described. This is a fully four-dimensional
quantum theory without an equivalent 3$+$1 formulation in terms of states on spacelike surfaces and
their unitary evolution between them. Rather, the usual 3$+$1 formulation is emergent for those
situations, and for those coarse grainings, where spacetime geometry behaves approximately 
classically.  
The intent of this development is not to propose a new quantum theory of gravity. This essentially low
energy theory suffers from the usual ultraviolet difficulties.  Rather, it is to employ this
theory as a model to discuss how quantum mechanics can be generalized to deal with quantum
geometry.

A common expectation is that spacetime is itself emergent from something more fundamental. In that
case a generalization of usual quantum mechanics will surely be needed and generalized quantum
theory can provide a framework for discovering it (Section 8). Emergence in quantum theory is 
discussed generally in Section 9.  Section 10 describes the emergence of Lorentz signatured
classical spacetimes from the no-boundary quantum state of the universe.

Section 11 concludes with some thoughts about whether quantum mechanics itself could be emergent
from something deeper. But before starting on the path of extending quantum theory so far we first
offer some remarks on where it is today in Section 2.

\setcounter{equation}{0}
\section{Quantum Mechanics Today}

Three features of quantum theory are striking from the present perspective: its
success, its rejection by some of our deepest thinkers, and the absence of compelling
alternatives. 

Quantum mechanics must be counted as one of the most successful of all physical theories.
Within the framework it provides, a truly vast range of phenomena can
be understood and that understanding is confirmed by precision experiment.  We perhaps have
little evidence for peculiarly quantum phenomena on large and even familiar scales, but
there is {\it no} evidence that all the phenomena we do see, from the smallest scales to the
largest of the universe, cannot be described in quantum mechanical terms and explained by
quantum mechanical laws. Indeed, the frontier to which quantum interference is confirmed
experimentally is advancing to ever larger, more `macroscopic' systems\footnote{For an
insightful and lucid review see \cite{Leg02}.}. The textbook electron two-slit experiment
shown schematically in Fig.~1 has been realized in the laboratory \cite{Ton89}. Interference
has been confirmed for the biomolecule tetraphenylporphyrin (C$_{44}$H$_{30}$N$_4$) and the
flurofullerine (C$_{60}$F$_{48}$) in analogous experiments \cite{Zei02} (Figure 2).  Experiments with
superconducting squids have demonstrated the coherent superposition of macroscopic currents
\cite{Moo00, Moo03, Fri03}.  In particular, the experiment of Friedman, \etal \cite{Fri03}
exhibited the coherent superposition of two circulating currents whose magnetic moments were
of order $10^{10} \mu_B$ (where $\mu_B=e\hbar/2m_ec$ is the Bohr magneton). 
Experiments under development will extend the boundary further \cite{Mar03}. 
Experiments of increasing ingenuity and sophistication have extended the regime in which
quantum mechanics has been tested. No limit to its validity has yet emerged.

\begin{figure}[t]
\parbox{2.3in}{\epsfig{file=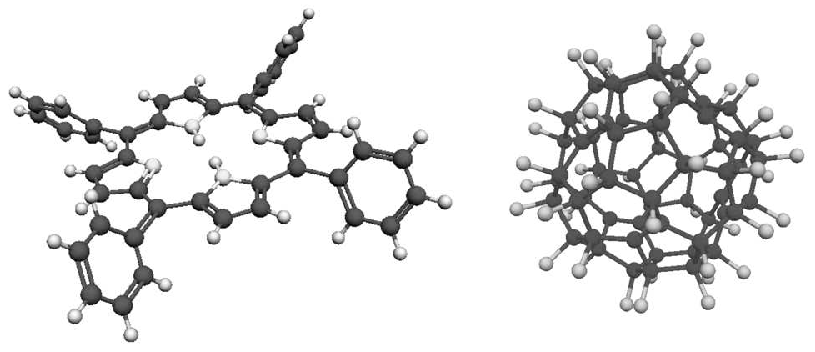,
width=2.3in,bbllx=5,bblly=0,bburx=150,bbury=100,clip}}
\hfill
\parbox{2.3in}{\rightline{\epsfxsize=2.5in \epsfbox{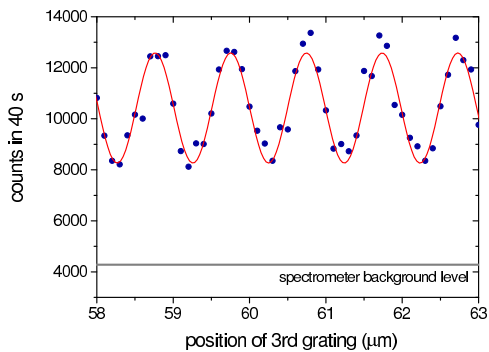}}}
\caption{ \small Interference of Biomolecules. The molecule tetraphenylporphyrin
(C$_{44}$H$_{30}$N$_{4}$) is shown at left.  Its quantum interference
fringes in a Talbot-Lau interferometer are shown at right from experiment carried
out in Anton Zeilinger's group (Hackerm{\" u}ller, {\it et al.}~\cite{Zei02}.) }
\end{figure}

Even while acknowledging its undoubted empirical success, many of our greatest minds have
rejected quantum mechanics as a framework for fundamental theory. Among the pioneers, the
names of Einstein, Schr\"odinger, DeBroglie, and Bohm stand out in this regard. Among our
distinguished contemporaries, Adler, Leggett, Penrose, and 't Hooft could probably be counted
in this category. Much of this thought has in common the intuition that quantum mechanics is
an effective approximation of a more fundamental theory built on a notion of reality closer
to that classical physics. 

Remarkably, despite eighty years of unease with its basic premises, and despite having been
tested only in a limited, largely microscopic, domain, no fully satisfactory alternative to
quantum theory has emerged.  By fully satisfactory we mean not only consistent with existing
experiment, but also incorporating other seemingly secure parts of modern physics
such as special relativity, field theory, and the standard model of elementary particle
interactions. As Steve Weinberg summarized the situation, ``It is striking that it has not so far been
possible to find a logically consistent theory that is close to quantum mechanics other than
quantum mechanics itself'' \cite{Wei92}. Alternatives to quantum theory meeting the above
criteria would be of great interest if only to guide experiment.

There are several directions under investigation today which aim at a theory from which
quantum mechanics would be emergent. Neither space nor the author's competence permit an
extensive discussion of these ideas. But we can mention some of the more important
ones.\footnote{The references to these ideas are obviously not exhaustive, nor are they
necessarily current. Rather, they are to typical sources. For an encyclopedic survey of
different interpretations and alternatives to quantum mechanics, see \cite{Aul00}.} 

Bohmian mechanics \cite{BH93} in its most representative form is a deterministic but highly
non-classical theory of particle dynamics whose statistical predictions largely coincide
with quantum theory \cite{Har04}. Fundamental noise \cite{Per98} or spontaneous dynamical
collapse of the wave function \cite{BG03, DH04} are the underlying ideas of another class of
model theories whose predictions are distinguishable from those of quantum theory, in principle.  
Steve Adler has proposed a statistical mechanics of deterministic matrix models from
which quantum mechanics is emergent \cite{Adl04}.  Gerard 't Hooft has a different set of
ideas for a determinism beneath quantum mechanism that are explained in his article in this
volume \cite{tHo06}. Roger Penrose has championed a role for gravity in state
vector reduction \cite{Pen00, Pen04}.  This has not yet developed into a detailed alternative
theory, but has suggested experimental situations in which the decay of quantum
superpositions could be observed \cite{Pen04, Mar03}.

In the face of an increasing domain of confirmed predictions of quantum theory and the 
absence as yet of compelling alternatives, it seems natural to extend
quantum theory as far as it will go --- to the largest scales of the universe and the
smallest of quantum gravity. That is the course we shall follow in this paper. But as
mentioned in the introduction, usual quantum theory must be generalized to apply to
cosmology and quantum spacetime. We amplify on the reasons in the next
section.

\setcounter{equation}{0}
\section{Spacetime and Quantum Theory}

Usual, textbook quantum theory incorporates definite assumptions about the nature of space
and time.  These assumptions are readily evident in the two laws of evolution for the
quantum state $\Psi$. The Schr\"odinger equation describes its unitary evolution between
measurements.
\begin{equation}
i\hbar\ \frac{\partial\Psi}{\partial t} = H\Psi\ .
\label{threeone}
\end{equation}
At the time of an ideal measurement, the state is projected on the outcome and renormalized
\begin{equation}
\Psi \to \frac{P\Psi}{\| P\Psi \|}\ .
\label{threetwo}
\end{equation}

\begin{table}
\begin{tabular}{|c|l|l|}
\multicolumn{3}{c}{\large\bf A Short History of Spacetime and Quantum Theory}\\\hline
\parbox[t]{1.5in}{Newtonian Physics} &
\parbox[t]{1.5in}{Fixed 3-d space and  \\a single universal \\ time $t$.} &
\parbox[t]{2.8in}{{\bf Non-relativistic Quantum Theory:}\\ The Schr\"odinger 
equation
\vskip .05in \centerline{ $
i \hbar (\partial \Psi/\partial t) = H \Psi
$} \vskip .04in
holds between measurements in the \\ Newtonian time $t$.} 
 \\ && \\ \hline
\parbox[t]{1.5in}{Special  Relativity} &
\parbox[t] {1.5in}{Fixed flat, 4-d\\ spacetime  with many\\ diifferent
 timelike\\ directions.} &
\parbox[t]{2.8in}{{\bf Relativistic Quantum Field\\  Theory:}\\
Choose a Lorentz frame with time $t$.\\  Then (between measurements) 
\vskip .05in \centerline{$
i \hbar (\partial \Psi/\partial t) = H \Psi\, .
$} \vskip .04in 
The results are unitarily equivalent to\\ those from any other choice of Lorentz\\
frame}
\\ && \\ \hline
\parbox[t]{1.5in}{General  Relativity}&
\parbox[t]{1.5in}{Fixed,  but curved \\ spacetime  geometry} &
\parbox[t]{2.8in}{{\bf Quantum Field Theory in Curved \\ Spacetime:}\\ Choose a foliating
famliy of spacelike\\ surfaces labeled by $t$.  Then  (between \\ measurements)
\vskip .05in \centerline{$
i \hbar (\partial \Psi/ \partial t) = H \Psi\, .
$} \vskip .04in
But the results are {\it not} generally\\ unitarily equivalent to other choices.}
\\ && \\ \hline
\parbox[t]{1.5in}{Quantum  Gravity}&
\parbox[t]{1.5in}{Geometry is {\it not}\\  fixed,  but rather a\\ quantum  variable}&
\parbox[t]{2.8in}{{\bf The Problem of Time:}\\ What replaces the Schr\"odinger\\
equation when there is no fixed\\ notion of time(s)?}
\\ && \\ \hline
\parbox[t]{1.5in}{M-theory, Loop \\ quantum gravity, \\ Posets, etc.}&
\parbox[t]{1.5in}{Spacetime is not even\\ a  fundamental variable}&
\parbox[t]{2.5in}{\vskip .15in \centerline{\Huge\bf ?}}
\\ &&
\\ \hline
\end{tabular}
\end{table}
\vspace{5mm}

The Schr\"odinger equation \eqref{threeone} assumes a fixed notion of time.  
In the non-relativistic theory,
$t$ is the absolute time of Newtonian mechanics. In the flat spacetime of special
relativity,
it is the time of any Lorentz frame. Thus, there are many times but results obtained in different
Lorentz frames, are unitarily equivalent. 

The projection in the second law of evolution
\eqref{threetwo} is in Hilbert space. But in field theory or particle mechanics, the Hilbert
space is constructed from configurations of fields or position in physical {\it space}. In
that sense it is the state on a spacelike surface that is projected \eqref{threetwo}. 

Because quantum theory incorporates notions of space and time, it has changed as our ideas
of space and time have evolved. The accompanying table briefly summarizes this co-evolution. It is possible
to view this evolution as a process of increasing generalization of the concepts in the
usual theory.  Certainly the two laws of evolution \eqref{threeone} and \eqref{threetwo} have to be
generalized somehow if spacetime geometry is not fixed. One such generalization is offered
in this paper, but there have been many other ideas \cite{POT}. And if spacetime geometry
is emergent from some yet more fundamental description, we can certainly expect that a
further generalization --- free of any reference to spacetime --- will be needed to describe
that emergence. The rest of this article is concerned with these generalizations.

\setcounter{equation}{0}
\section{The Quantum Mechanics of Closed\\
 Systems}

This section reviews, very briefly, the elements of the modern quantum mechanics of closed
systems\footnote{See, {\it e.g.} \cite{Gri02,Omn94,Gel94} for by now classic expositions at
length or \cite{Har93a} for a shorter summary.}  aimed at a quantum mechanics for cosmology.
To keep the present discussion manageable we focus on a simple model universe of
particles moving in a very large box (say $\gtwid$ 20,000 Mpc in linear
dimension). Everything is contained within the box, in particular galaxies, stars, planets,
observers and observed (if any), measured subsystems, and the apparatus that measures them. 

We assume a fixed background spacetime supplying well-defined notions of time.  The usual
apparatus of Hilbert space, states, operators, Feynman paths, etc. can then be employed in a
quantum description of the contents of the box. The essential theoretical inputs to the
process of prediction are the Hamiltonian $H$ and the initial quantum state $|\Psi\rangle$
(the `wave function of the universe'). These are assumed to be fixed and given.

The most general objective of a quantum theory for the box is the prediction of the
probabilities of exhaustive sets of coarse-grained alternative time histories of the
particles in the closed system.  For instance, we might be interested in the
probabilities of an alternative set of histories describing the progress of the Earth around
the Sun. Histories of interest here are typically very coarse-grained for at least three reasons:
They deal with the position of the Earth's center-of-mass and not with the positions
of all the particles in the universe.  The center-of-mass position is not specified
to arbitrary accuracy, but to the error we might observe it. The center-of-mass
position is not specified at all times, but typically at a series of times.

But, as described in the Introduction, not every set of alternative histories that may be
described can be assigned consistent probabilities because of quantum interference.  Any
quantum theory must therefore not only specify the sets of alternative coarse-grained histories, 
but also give a rule identifying which sets of histories can be consistently assigned probabilities 
as well as what those probabilities are.   In the quantum mechanics of closed systems, 
that rule is simple: probabilities can be assigned to just those sets of histories for which the 
quantum interference between its members is negligible as a consequence of the Hamiltonian $H$ and the
initial state $|\Psi\rangle$. We now make this specific for our model universe of particles
in a box.  

Three elements specify this quantum theory.  To facilitate later discussion, we
give these in a spacetime sum-over-histories formulation.

\begin{enumerate}

\item {\sl Fine-grained histories}: The most refined description of the particles from the
initial time $t=0$ to a suitably large final time $t=T$ gives their position at all times in
between, {\it i.e.}~their Feynman paths.  We denote these simply by $x(t)$.

\item {\sl Coarse-graining}: The general notion of coarse-graining is a partition of the
fine-grained paths into an exhaustive set of mutually exclusive classes $\{c_\alpha\},
\alpha=1, 2, \cdots$. For instance, we might partition the fine-grained histories of the
center-of-mass of the Earth by which of an exhaustive and exclusive set of position intervals
$\{\Delta_\alpha\}, \alpha=1, 2, \cdots$ the center-of-mass passes through at a series of
times $t_1, \cdots t_n$. Each coarse-grained history consists of the bundle of fine-grained
paths that pass through a specified sequence of intervals at the series of times.  Each
coarse-grained history specifies an orbit where the center-of-mass position is localized to a
certain accuracy at a sequence of times.

\item{\sl Measure of Interference}: Branch state vectors $|\Psi_\alpha\rangle$ can be defined for
each coarse-grained history in a partition of the fine-grained histories into classes
$\{c_\alpha\}$ as follows
\begin{equation}
\langle x |\Psi_\alpha\rangle = \int_{c_\alpha} \delta x\ \exp (iS[x(t)]/\hbar)
\, \langle x^\prime |\Psi\rangle\, .
\label{fourone}
\end{equation}
Here, $S[x(t)]$ is the action for the Hamiltonian $H$. The integral is over all paths
starting at $x^\prime$ at $t=0$, ending at $x$ at $t=T$, and contained 
in the class $c_\alpha$. This
includes an integral over $x^\prime$. (For those preferring the Heisenberg picture, this is
equivalently
\begin{equation}
|\Psi_\alpha\rangle = e^{-iHT/\hbar} P^n_{\alpha_n} (t_n) \cdots P^1_{\alpha_1} (t_1)\,
|\Psi\rangle
\label{fourtwo}
\end{equation}
when the class consists of restrictions to position intervals
at a series of times and the $P$'s are the projection operators representing them.)

The measure of quantum interference between two coarse-grained histories is the overlap of
their branch state vectors
\begin{equation}
D(\alpha^\prime, \alpha) \equiv \langle \Psi_{\alpha^\prime} |\Psi_\alpha\rangle\, .
\label{fourthree}
\end{equation}
This is called the {\sl decoherence functional}. 

\end{enumerate}

When the interference between each pair of histories in a coarse-grained set is negligible
\begin{equation}
\langle\Psi_\alpha |\Psi_\beta\rangle \approx 0\ {\rm all}\ \alpha\not=\beta\, ,
\label{fourfour}
\end{equation}
the set of histories is said to {\it decohere}\footnote{This is the {\it medium} decoherence
condition. For a discussion of other conditions, see, {\it e.g.} \cite{GH90b, GH95, Har04a}.}. 
The probability of an individual history in a decoherent set is
\begin{equation}
p(\alpha)=\|\, |\Psi_\alpha\rangle\|^2\, .
\label{fourfive}
\end{equation}
The decoherence condition \eqref{fourthree} is a sufficient condition for the probabilities
\eqref{fourfour} to be consistent with the rules of probability theory. Specifically, the
$p$'s obey the sum rules
\begin{equation}
p(\bar\alpha) \approx \sum_{\alpha\in\bar\alpha} p(\alpha)
\label{foursix}
\end{equation}
where $\{\bar c_{\bar\alpha}\}$ is any coarse-graining of the set $\{c_\alpha\}$, {\it i.e.}~a
further partition into coarser classes.  It was the failure of such a sum rule that prevented
consistent probabilities from being assigned to the two histories  previously discussed in the 
two-slit experiment (Figure 1). That set of histories does not decohere.

Decoherence of familiar quasiclassical variables is widespread in the universe. Imagine, for
instance, a dust grain in a superposition of two positions, a multimeter apart, deep in
intergalactic space. The $10^{11}$ cosmic background photons that scatter off the dust grain
every second dissipate the phase coherence between the branches corresponding to the two
locations on the time scale of about a nanosecond \cite{JZ85}.

Measurements and observers play no fundamental role in this generalization of usual quantum
theory. The probabilities of measured outcomes can, of course, be computed and are given to
an excellent approximation by the usual story.\footnote{See, {\it e.g.}~\cite{Har91a},
Section II.10.} But, in a set of histories where they
decohere, probabilities can be assigned to the position of the Moon when it is not being
observed and to the values of density fluctuations in the early universe when there were
neither measurements taking place nor observers to carry them out.

\setcounter{equation}{0}
\section{Quantum Theory in 3$+$1 Form}

The quantum theory of the model universe in a box in the previous section is in fully 4-dimensional
spacetime form. The fine-grained histories are paths in spacetime, the coarse-grainings were
partitions of these, and the measure of interference was constructed by spacetime path
integrals. No mention was made of states on spacelike surfaces or their unitary evolution.

However, as originally shown by Feynman \cite{Fey48, FH65}, this spacetime formulation is
equivalent to the familiar 3$+$1 formulation in terms of states on spacelike surfaces and
their unitary evolution through a foliating family of such surfaces.  This section briefly
sketches that equivalence emphasizing properties of spacetime and the fine-grained histories that
are necessary for it to hold.

\begin{figure}[t]
\centerline{\epsfxsize=5in \epsfbox{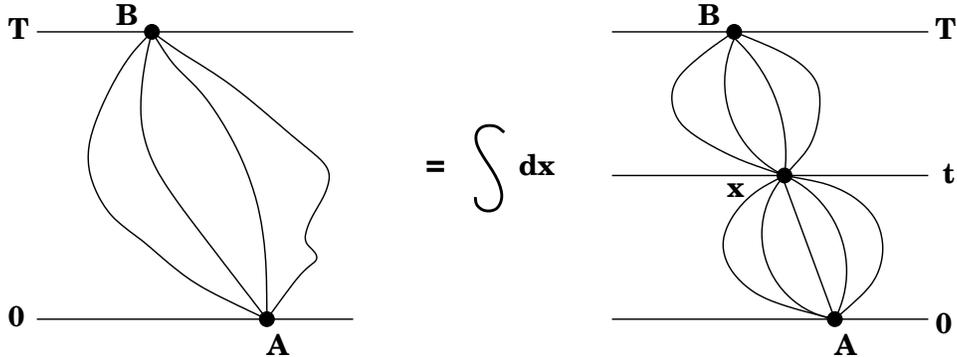}}
\caption{\small The origin of states on a spacelike surface.  These spacetime diagrams
are a schematic representation of Eq. \eqref{fiveone}. The amplitude for a particle
to pass from point $A$ at time $t=0$ to a point $B$ at $t=T$ is a sum over all
paths connecting them weighted by $\exp(i S[x(t)])$. That sum can be factored
across an intermediate constant time surface as shown at right into product of a
sum from $A$ to $x$ on the surface and a sum from $x$ to $B$ followed by a sum
over all $x$. The sums in the product define states on the surface of constant
time at $t$. The integral over $x$ defines the inner product between such states,
and the path integral construction guarantees their unitary evolution in $t$. Such
factorization is possible only if the paths are single valued functions of time. }
\end{figure}

The key observation is illustrated in Figure 3. Sums-over-histories that are single-valued in
time can be factored across constant time surfaces. A formula expressing this idea is
\begin{equation}
\int_{[A, B]} \delta x\, e^{iS[x(t)]/\hbar} = \int dx\, \psi^*_B (x, t)\psi_A (x, t)\, .
\label{fiveone}
\end{equation}
The sum on the left is over all paths from $A$ at $t=0$ to $B$ at $t=T$. The amplitude
$\psi_A(x, t)$ is the sum of $\exp\{iS[x(t)]\}$ over all paths from $A$ at $t=0$ to $x$ at
a time $t$ between 0 and $T$. The amplitude $\psi_B(x,t)$ is similarly constructed from the paths
between $x$ at $t$ to $B$ at $T$.

The wave function $\psi_A(x,t)$ defines a state on constant time surfaces.  Unitary evolution
by the Schr\"odinger equation follows from its path integral
construction.\footnote{Reduction of the state vector \eqref{threetwo} also follows from the path
integral construction \cite{Cav86} when histories are coarse-grained by intervals of
position at various times.} The inner product between states
defining a Hilbert space is specified by \eqref{fiveone}. In this way, the familiar 3$+$1
formulation of quantum mechanics is recovered from its spacetime form.

The equivalence represented in \eqref{fiveone} relies on several special assumptions
about the nature of spacetime and the fine-grained histories.  In particular, it
requires\footnote{The usual 3$+$1 formulation is also restricted to coarse-grained histories
specified by alternatives at definite moments of time. More general spacetime
coarse-grainings that are defined by quantities that extend over time can be used in the
spacetime formulation. (See, {\it e.g.}~\cite{HB05} and references therein.)  Spacetime alternatives are the only ones available in a diffeomorphism invariant quantum graviity. }: 

\begin{itemize}

\item A fixed Lorentzian spacetime geometry to define timelike and spacelike directions.

\item A foliating family of spacelike surfaces through which states can evolve.

\item Fine-grained histories that are single-valued in the time labeling the spacelike
surfaces in the foliating family.

\end{itemize}

As an illustrative example where the equivalence does not hold, consider quantum field 
theory in a fixed background spacetime
with closed timelike curves (CTCs) such as those that can occur in wormhole spacetimes
\cite{MTY88}. The fine-grained histories are four-dimensional field configurations that are
single-valued on spacetime.  But there is no foliating family of spacelike surfaces with
which to define the Hamiltonian evolution of a quantum state. Thus, there is no usual 3$+$1
formulation of the quantum mechanics of fields in spacetimes with CTCs.

However, there is a four-dimensional sum-over-histories formulation of field theory in
spacetimes with CTCs \cite{Har94a, FPS92, Ros98}. The resulting theory has some
unattractive properties such as acausality and non-unitarity. But it does illustrate how
closely usual quantum theory incorporates particular assumptions about spacetime, and also how these requirements can be relaxed in a suitable generalization of the usual
theory.

\setcounter{equation}{0}
\section{Generalized Quantum Theory}

In generalizing usual quantum mechanics to deal with quantum spacetime, some of its features
will have to be left behind and others retained. What are the minimal essential
features that characterize a quantum mechanical theory?  The generalized quantum
theory framework \cite{Har91a, Har93a, Ishsum95} provides one answer to this
question. Just three elements abstracted from the quantum mechanics of closed systems
in Section 4 define a generalized quantum theory.

\begin{itemize}

\item {\sl Fine-grained Histories}:  The sets of alternative fine-grained histories of
the closed system which are the most refined descriptions of it physically possible.

\item {\sl Coarse-grained Histories}: These are partitions of a set of
fine-grained histories into an exhaustive set of exclusive classes $\{c_\alpha\},
\alpha=1, 2 \cdots$. Each class is a coarse-grained history.

\item {\sl Decoherence Functional}: A measure of quantum interference $D(\alpha,
\alpha^\prime)$ between pairs of histories in a coarse-grained set, meeting the
following conditions:

\begin{itemize}

\item[i.] Hermiticity: $D(\alpha, \alpha^\prime)=D^* (\alpha^\prime, \alpha)$

\item[ii.] Positivity: $D(\alpha, \alpha) \geq 0$

\item[iii.] Normalization: $\Sigma_{\alpha\alpha^\prime} D(\alpha, \alpha^\prime)=1$

\item[iv.] Principle of superposition: If $\{\bar c_{\bar \alpha}\}$ is a further
coarse-graining of $\{c_\alpha\}$, then 
\[\bar D(\bar\alpha, \bar\alpha^\prime) =
\sum_{{\alpha\in\bar\alpha}\atop {\alpha^\prime\in\bar\alpha^\prime}}
D(\alpha,\alpha^\prime)
\]

\end{itemize}
\end{itemize}

Probabilities $p(\alpha)$ are assigned to sets of coarse-grained histories when they
decohere according to the basic relation
\begin{equation}
D(\alpha, \alpha^\prime)\approx \delta_{\alpha\alpha^\prime}\, p(\alpha)\, .
\label{sixone}
\end{equation}
These $p(\alpha)$ satisfy the basic requirements for probabilities as a consequence of
i)--iv) above. In particular, they satisfy the sum rule
\begin{equation}
p(\bar\alpha) = \sum_{\alpha\in\bar\alpha} p(\alpha)
\label{sixtwo}
\end{equation}
as a consequence of i)--iv) and decoherence. For instance, the probabilities of an exhaustive 
set of alternatives always sum to 1.

The sum-over-histories formulation of usual quantum mechanics given in Section 4 is
a particular example of a generalized quantum theory. The decoherence functional
\eqref{fourone} satisfies the requirements i)--iv). But its particular form is not
the only way of constructing a decoherence functional.  Therein lies the possibility
of generalization.

\setcounter{equation}{0}
\section{A Quantum Theory of Spacetime Geometry}

The low energy, effective theory of quantum gravity is a quantum version of general
relativity with a spacetime metric $g_{\alpha\beta}(x)$ coupled to matter fields.  Of course,
the divergences of this effective theory have to be regulated to extract predictions
from it.\footnote{Perhaps, most naturally by discrete approximations to geometry such
as the Regge calculus (see, {\it e.g.}~\cite{Har85a, HW04})}. These predictions 
can therefore be expected to be accurate only for limited
coarse-grainings and certain states. But this effective theory does supply an
instructive model for generalizations of quantum theory that can accommodate quantum
spacetime. This generalization is sketched in this section.

The key idea is that the fine-grained histories do not have to represent evolution
{\it in} spacetime. Rather they can be histories {\it of} spacetime. For this discussion 
we take these histories to be spatially closed cosmological four-geometries represented by metrics
$g_{\alpha\beta}(x)$ on a fixed manifold $M=\boldsymbol{R}\times M^3$ where $M^3$ is a closed
3-manifold. For simplicity, we restrict attention to a single scalar matter field $\phi(x)$. 

The three ingredients of a generalized quantum theory for spacetime geometry are then
as follows:

\begin{itemize}

\item {\sl Fine-grained Histories}: A fine-grained history is defined by a
four-dimensional metric and matter field configuration on $M$.

\item {\sl Coarse-grainings}: The allowed coarse-grainings are partitions of the metrics and
matter fields into four-dimensional {\sl diffeomorphism invariant} classes
$\{c_\alpha\}$.

\item {\sl Decoherence Functional}:  A decoherence functional constructed on
sum-over-history principles analogous to that described for usual quantum theory in
Section 4.  Schematically, branch state vectors $|\Psi_\alpha\rangle$ can be
constructed for each coarse-grained history by summing over the metrics and fields in
the corresponding class $c_\alpha$ of fine-grained histories, {\it viz.}
\begin{equation}
|\Psi_\alpha\rangle = \int_{c_\alpha} \delta g \delta\phi\, \exp\{iS[g,
\phi]/\hbar\}\, |\Psi\rangle\, .
\label{sevenone}
\end{equation}
A decoherence functional satisfying the requirements of Section 6 is
\begin{equation}
D(\alpha^\prime, \alpha)=\langle\Psi_{\alpha^\prime} | \Psi_\alpha\rangle\, .
\label{seventwo}
\end{equation}
Here, $S[g, \phi]$ is the action for general relativity coupled to the field
$\phi(x)$, and $|\Psi\rangle$ is the initial cosmological state. The construction is
only schematic because we did not spell out how the functional integrals are defined
or regulated, nor did we specify the product between states that is implicit in both
\eqref{sevenone} and \eqref{seventwo}. These details can be made specific in models
\cite{Har95c, HM97, HC04}, but they will not be needed for the subsequent discussion.

\end{itemize}

A few remarks about the coarse-grained histories may be helpful. To every physical
assertion that can be made about the geometry of the universe and the fields within,  
there corresponds
a diffeomorphism invariant partition of the fine-grained histories into the class
where the assertion is true and the class where it is false. The notion of
coarse-grained history described above therefore supplies the most general notion of
alternative describable in spacetime form.  Among these we do not expect to find
local alternatives because there is no diffeomorphism invariant notion of locality. In 
particular, we do not expect to find alternatives specified at a moment of time.  We do  
expect to find alternatives referring to the kind of relational observables discussed in
\cite{GMH06} and the references therein. We also expect to find observables referring
to global properties of the universe such as the maximum size achieved over the history
of its expansion.

This generalized quantum mechanics of spacetime geometry is in fully spacetime form with
alternatives described by partitions of four-dimensional histories and a decoherence
functional defined by sums over those histories. It is analogous to the spacetime formulation
of usual quantum theory reviewed in Section 4.

However, unlike the theory in Section 4, we cannot expect an equivalent 3$+$1 formulation,
of the kind described in Section 5, 
expressed in terms of states on spacelike surfaces and their unitary evolution between these
surfaces.  The fine-grained histories are not `single-valued' in any
geometrically defined variable labeling a spacelike surface. They therefore cannot be
factored across a spacelike surface as in \eqref{fiveone}. More precisely, there is no
geometrical variable that picks out a unique spacelike surface in all
geometries.\footnote{Spacelike surfaces labeled by the trace of the extrinsic curvature $K$
foliate certain classes of classical spacetimes  obeying the Einstein equation \cite{MT80}. However, there is no reason to
require that non-classical histories be foliable in this way. It is easy to construct
geometries where surfaces of a given $K$ occur arbitrarily often.} 

Even without a unitary evolution of states the generalized quantum theory is fully
predictive because it assigns probabilities to the most general sets of coarse-grained alternative
histories described in spacetime terms when these are decoherent.

How then is usual quantum theory used every day, with its unitarily
evolving states, connected to this generalized quantum theory that is free from them? The answer is
that usual quantum theory is an approximation to the more general framework that is
appropriate for those coarse-grainings and initial state $|\Psi\rangle$ for which spacetime
behaves classically. One equation will show the origin of this relation. Suppose we have a
coarse-graining that distinguishes between fine-grained geometries only by their behavior on
scales well above the Planck scale. Then, for suitable states $|\Psi\rangle$ we expect that
the integral over metrics in \eqref{seventwo} can be well approximated semiclassically by the
method of steepest descents. Suppose further for simplicity that only a single classical
geometry with metric $\hat g_{\alpha\beta}$ dominates the semiclassical approximation. Then,
\eqref{seventwo} becomes
\begin{equation}
|\Psi_\alpha\rangle \approx \int_{\hat c_\alpha} \delta\phi\, \exp\{iS[\hat g,
\phi]/\hbar\}\, |\Psi\rangle
\label{seventhree}
\end{equation}
where $\hat c_\alpha$ is the coarse-graining of $\phi(x)$ arising from $c_\alpha$ and the
restriction of $g_{\alpha\beta}(x)$ to $\hat g_{\alpha\beta}(x)$. Eq.~\eqref{seventhree}
effectively defines a quantum theory of the field $\phi(x)$ in the fixed background spacetime
with the geometry specified by $\hat g_{\alpha\beta}(x)$. This is familiar territory. Field
histories are single valued on spacetime. Sums-over-fields can thus be factored across
spacelike surfaces in the geometry $\hat g$ as in \eqref{fiveone} to define field states on 
spacelike surfaces, their
unitary evolution, and their Hilbert space product. Usual quantum theory is thus recovered
when spacetime behaves classically and provides the fixed spacetime geometry on which usual
quantum theory relies.

From this perspective, familiar quantum theory and its unitary evolution of states is an
effective approximation to a more general sum-over-histories formulation of quantum theory.
The approximation is appropriate for those coarse-grainings and initial states in which
spacetime geometry behaves classically.

\setcounter{equation}{0}
\section{Beyond Spacetime}

The generalized quantum theory of spacetime sketched in the previous section assumed that
geometry was a fundamental variable --- part of the description of the fine-grained
histories. But on almost every frontier in quantum gravity one finds the idea that continuum
geometry is not fundamental, but will be replaced by something more fundamental. This is true
for string theory \cite{Sei06}, loop quantum gravity \cite{AL04},  and the causal set program \cite{Dow05, HS06}
although space does not permit a review of these speculations.

Can generalized quantum theory serve as a framework for theories where spacetime is emergent
rather than fundamental? Certainly we cannot expect to have a notion of `history'. But we can
expect some fine-grained description, or a family of equivalent ones, and that is enough. A
generalized quantum theory needs

\begin{itemize}

\item The possible fine-grained descriptions of the system.

\item The coarse-grained descriptions constructed from the fine-grained ones.

\item A measure of quantum interference between different coarse-grained descriptions
respecting conditions i)--iv) in Section VI.

\end{itemize}
Generalized quantum theory requires neither space nor time and can therefore serve as the 
basis for a quantum theory in which spacetime is emergent.

\setcounter{equation}{0}
\section{Emergence/Excess Baggage}

The word `emergent' appears in a number of places in the previous discussion. It 
probably has many meanings.  This section aims at a more precise understanding of what is 
meant by the term in this essay.

Suppose we have a quantum theory defined by certain sets of fine-grained histories,
coarse-grainings, and a decoherence functional.  Let's call this the 
{\it fundamental} theory. It may happen that the decoherence and probabilities of limited 
kinds of sets of
coarse-grained histories are given approximately by a second, {\it effective} theory. The two theories are related in
the following way:

\begin{itemize}

\item Every fine-grained history of the effective theory is a coarse-grained history of the
fundamental theory.

\item The decoherence functionals approximately agree on a limited class of sets of
coarse-grained histories.
\begin{equation}
D^{\rm fund} (\alpha^\prime, \alpha) \approx D^{\rm eff} (\alpha^\prime, \alpha)\, .
\label{nineone}
\end{equation}
On the right, $\alpha^\prime$ and $\alpha$ refer to the fine-grained histories of the
effective theory.  On the left, they refer to the corresponding coarse-grained histories of
the fundamental theory.

\end{itemize}
When two theories are related in this way we can say that the effective theory is {\it
emergent} from the fundamental theory. Loosely we can say that the restrictions, and the
concepts that characterize them, are emergent. It should be emphasized that an approximate 
equality like \eqref{nineone} can be expected to hold, not just as a consequence of the
particular dynamics incorporated into decoherence functionals, but also only for particular 
states.

Several examples of emergence in this sense have been considered in this essay: There is the
possible emergence of a generalized quantum theory of spacetime geometry from a theory in which
spacetime is not fundamental.  There is the emergence of a 3$+$1 quantum theory of fields in a
fixed background geometry from a four-dimensional generalized quantum theory in which geometry is
a quantum variable.  There is the emergence of the approximate quantum mechanics of measured
subsystems (textbook quantum theory) from the quantum mechanics of the universe.  And there is
the emergence of classical physics from quantum phyaics.

Instead of looking at an effective theory as a restriction of a more fundamental one, we may
look at the fundamental theory as a generalization of the effective one.  That perspective is
important because generalization is a way of searching for more comprehensive theories of
nature. In passing from the specific to the more general some ideas have to be discarded.
They are often ideas that were once perceived to be general because of our special place in
the universe and the limited range of our experience.  But, in fact, they arise from special
situations in a more general theory. They are `excess baggage' that has to be discarded to
reach a more comprehensive theory \cite{Har90b}. Emergence and excess baggage are two ways of
looking at the same thing.
\begin{table}[p]
\thicklines
\setlength{\unitlength}{.41cm}
\begin{picture}(39,56)
\put(34.2,14){\line(0,1){37}}
\put(31,52){\makebox(6,3){\parbox{6\unitlength}{\begin{center} \large\bf Discarding\\
Excess\\ Baggage \end{center}}}}
\put(31,12){\makebox(6,3){\parbox{6\unitlength}{\begin{center} \large \bf Emergence
\end{center}}}}
\put(34.2,12){\vector(0,-1){6}}
\put(34.2,56){\vector(0,1){6}}

\put(0,56){\flbox{ {\Huge \bf ?} \\ \bf Spacetime not \\ Fundamental}}
\put(6,52){\dln}
\put(6,54){\hln}
\put(0,46){\flbox{\bf QM of Closed Systems: \\ Quantum spacetime \\ Quantum matter}}
\put(6,42){\dln}
\put(6,44){\hln}
\put(0,36){\flbox{\bf QM of Closed Systems: \\ Classical spacetime \\ Quantum matter}}
\put(6,32){\dln}
\put(6,34){\hln}
\put(0,26){\flbox{\bf Approximate QM \\ of Measured Subsystems\\ (textbook QM)}}
\put(6,22){\dln}
\put(6,24){\hln}
\put(0,16){\flbox{\bf Classical Physics}}
\put(6,12){\dln}
\put(6,14){\hln}
\put(0,6){\flbox{\bf Specific Systems: \\ stars,  planets \\ biological species, etc}}
\put(16,51){\flboxa{\bf Spacetime and \\ Histories}}
\put(16,41){\flboxa{\bf States on \\ spacelike  surfaces \\ and their \\ unitary evolution}}
\put(16,31){\flboxa{\bf Quasiclassical realm,  \\ measurements \\ as fundamental}}
\put(16,21){\flboxa{ \bf Determinism}}
\put(16,11){\flboxa{\bf Specific \\ Regularities}}

\end{picture}
\end{table}

Physics is replete with examples of emergence and excess baggage ranging from Earth-centered
theories of the solar system to quantum electrodynamics. The chart on the
next page helps understand the stages of emergence and generalization in quantum mechanics
discussed in this essay provided it is not taken too rigidly or without qualification.

The chart can be read in two ways: Reading from the bottom up, the boxes on the left describe a
path of generalization --- from the specific to the general. Starting from the regularities
of specific systems such as the planetary orbits, we move up to the general laws of classical
physics, to textbook quantum theory, through various stages of assumptions about spacetime,
to a yet unknown theory where spacetime is not fundamental. The excess baggage that must be
jettisoned at each stage to reach a more general perspective is indicated in the middle tower
of boxes.

Reading from the top down the chart tells a story of emergence. Each box on the left stands in
the relation of an effective theory to the one before it. The middle boxes now describe
phenomena that are emergent at each stage.

\setcounter{equation}{0}
\section{Emergence of Signature}

Classical spacetime has Lorentz signature. At each point it is possible to choose one
timelike direction and three orthogonal spacelike ones. There are no physical spacetimes with
zero timelike directions or with {\it two} timelike directions. But is such a seemingly basic
property fundamental, or is it rather, emergent from a quantum theory of spacetime which
allows for all possible signatures? This section sketches a simple model where that happens.

Classical behavior requires particular states \cite{Har94b}. Let's consider the possible
classical behaviors of cosmological geometry assuming the `no-boundary' quantum state of the
universe \cite{HH83} in a theory with only gravity and a cosmological constant $\Lambda$. 
The no-boundary wave function is given by a sum-over-geometries of the schematic form
\begin{equation}
\Psi\, [h] = \int_e \delta g\, e^{-I\, [g]/\hbar}\, .
\label{tenone}
\end{equation}
For simplicity, we consider a fixed manifold\footnote{Even the notion of manifold may be 
emergent in a more general theory of certain complexes \cite{Har85b, SW93}.} $M$.
The key requirement is that it be compact
with one boundary for the argument of the wave function and no other boundary.  The
functional $I\, [g]$ is the Euclidean action for metric defining the geometry on $M$. 
The sum is over a complex contour ${\cal C}$ of $g$'s  that have finite action and match the three-metric $h$ on the
boundary that is  the argument of $\Psi$.

Quantum theory predicts classical behavior when it predicts high probability for histories 
exhibiting the correlations in time implied by classical
deterministic laws \cite{GH93a, Har94b}. The state $\Psi$ is an input to the process of
predicting those probabilities as described in Section 7. However, plausibly the output for
the predicted classical spacetimes in this model are the extrema of the action in 
\eqref{tenone}. We will assume
this (see \cite{Har95c} for some justification). Further, to keep the discussion
manageable, we will restrict it to the {\it real} extrema.  These are the real tunneling
geometries discussed in a much wider context in \cite{GH90}.

\begin{figure}[t]
\centerline{\epsfig{height=2.5in, file=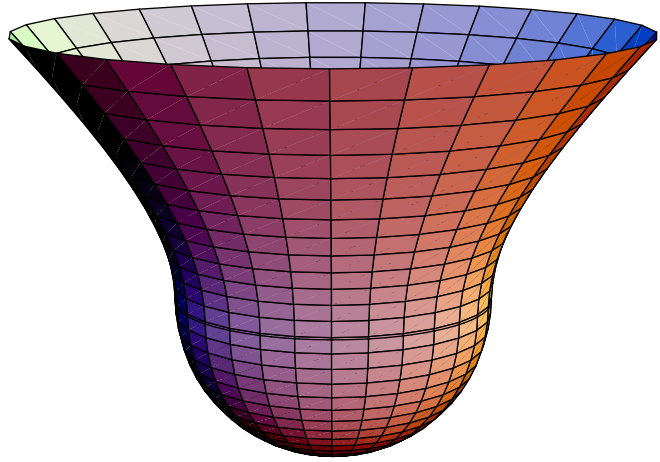, bbllx=88, bblly=60,
bburx=376, bbury=210, clip}}
\caption{\small The emergence of the Lorentz signature $(-,+,+,+)$ of spacetime. The
semiclassical  geometry describing  a  classical spacetime which becomes large
according to the `no-boundary' proposal for the universe's quantum state. The
model is pure gravity and a cosmological constant.   Purely Euclidean geometries
$(+,+,+,+)$  or purely Lorentzian geometries are not allowed as described in the text.
What is allowed is the
real tunneling geometry illustrated above consisting of half a Euclidean
four-sphere joined smoothly onto an expanding Lorentzian de Sitter space at the
moment of maximum contraction. This can be described as the nucleation of
classical Lorentz signatured spacetime. There is no similar nucleation of a
classical geometry with signature $(-,-,+,+)$ because it could not match the
Euclidean one across a spacelike surface. }
\end{figure}

Let us ask for the semiclassical geometries which become large, {\it i.e.}~contain symmetric three
surfaces with size much larger than $(1/\Lambda)^{1/2}$. There are none with Euclidean
signature. The purely Euclidean extremum is the round four-sphere with linear
size $(1/\Lambda)^{1/2}$ and contains no symmetric three surfaces
with larger size.
There are none with purely Lorentzian signature either because these
cannot be regular on $M$. There are, however, tunneling solutions of the kind illustrated
in Figure 4 in which half of a Euclidean four-sphere is matched to expanding DeSitter space
across a surface of vanishing extrinsic curvature.  

Could a spacetime with two time and two space directions be nucleated in this way? The
answer is `no' because the geometry on a surface could not have the three spacelike
directions necessary to match onto the half of a four-sphere.

Thus, in this very simple model, with many assumptions, if we live in a large universe it
must have one time and three space dimensions. The Lorentzian signature of classical
spacetime is an emergent property from an underlying theory not committed to this
signature.

\setcounter{equation}{0}
\section{Beyond Quantum Theory}

The path of generalization in the previous sections began with the textbook quantum
mechanics of measurement outcomes in a fixed spacetime and ended in a quantum
theory where neither measurements nor spacetime are fundamental. In this journey, the
principles of generalized quantum theory are preserved, in particular the idea of quantum
interference and the linearity inherent in the principle of superposition. But the end of this
path is strikingly different from its beginning.

The founders of quantum theory thought that the indeterminacy of quantum theory
``reflected the unavoidable interference in measurement dictated by the magnitude of the
quantum of the action'' (Bohr). But what then is the origin of quantum indeterminacy in a
closed quantum universe which is never measured? Why enforce the principle of
superposition in a framework for prediction of the universe which has but a single quantum
state? In short, the endpoint of this journey of generalization forces us to ask John
Wheeler's famous question, ``How come the quantum?'' \cite{Whe86}.

Could quantum theory itself be an emergent effective theory? Many have thought so (Section
2). Extending quantum mechanics until it breaks could be one route to finding out. 
`Traveler, there are no paths, paths are made by walking.'

\section{Conclusion}

Does quantum mechanics apply to spacetime?  The answer is `yes' provided that its familiar
textbook formulation is suitably generalized. It must be generalized in two directions.
First, to a quantum mechanics of closed systems, free from a fundamental role for
measurements and observers and therefore applicable to cosmology.  Second, it must be
generalized so that it is free from any assumption of a fixed spacetime geometry and therefore
applicable when spacetime geometry is a quantum variable.

Generalized quantum theory built on the pillars of fine-grained histories, coarse-graining,
and decoherence provides a framework for investigating such generalizations.  The fully,
four-dimensional sum-over-histories effective quantum theory of spacetime geometry sketched
in Section 7 is one example.  In such fully four-dimensional generalizations of the usual
theory, we cannot expect to recover an equivalent 3$+$1 formulation in terms of the unitary
evolution of states on spacelike surfaces. There is no fixed notion of spacelike surface.
Rather, the usual 3$+$1 formulation emerges as an effective approximation to the more
general story for those coarse grainings and initial states in which spacetime geometry
behaves classically.

If spacetime geometry is not fundamental, quantum mechanics will need further generalization
and generalized quantum theory provides one framework for exploring that.

\vskip .26 in

\noindent {\bf Acknowledgments}:

The author is grateful to Murray Gell-Mann and David Gross for delivering this paper at the
Solvay meeting when he was unable to do so. Thanks are due to Murray Gell-Mann for
discussions and collaboration on these issues over many years. Preparation of this report
was supported, in part, by NSF Grant PHY02-44764

\vspace{5mm}




\end{document}